\documentclass[12pt]{iopart}
\usepackage{epsf}

\newcommand{\ket}[1]{|#1\rangle}
\newcommand{\on}[0]{|1\rangle\!\langle 1|}
\newcommand{\an}[0]{\hat{b}_\textbf{k}}
\newcommand{\cre}[0]{\hat{b}^\dag_\textbf{k}}

\newcommand{\sumk}[0]{\sum_\textbf{k}}
\newcommand{\kx}[0]{\textbf{k}\cdot\textbf{x}}

\newcommand{\ak}[0]{A_k}
\newcommand{\bv}[1]{\textbf{#1}}
\newcommand{\bx}[0]{\textbf{x}}
\newcommand{\dx}[0]{\rmd^{D}\textbf{x}}

\newcommand{\pp}[0]{\hat{\phi}}
\newcommand{\ppc}[0]{\hat{\phi}_{\sigma}}

\newcommand{\nn}[0]{\hat{n}}
\newcommand{\x}[0]{(\bx)}
\newcommand{\xt}[0]{(\bx,t)}
\newcommand{\xot}[0]{(\bx_0,t)}

\newcommand{\vol}[1]{\frac{#1}{\sqrt{2L^D}}}
\newcommand{\wk}[0]{\omega_k}
\newcommand{\pex}[0]{\exp{[\rmi\kx]}}
\newcommand{\nex}[0]{\exp{[-\rmi\kx]}}
\newcommand{\hc}[0]{\mathrm{h.c.}\;}

\begin{document}

\title{Probing BEC phase fluctuations with atomic quantum dots}

\author{M~Bruderer and D~Jaksch}
\address{Clarendon Laboratory, University of Oxford, Parks
Road, Oxford OX1 3PU, United Kingdom}

\ead{m.bruderer@physics.ox.ac.uk}

\begin{abstract}
We consider the dephasing of two internal states $\ket{0}$ and
$\ket{1}$ of a trapped impurity atom, a so-called atomic quantum
dot (AQD), where only state $\ket{1}$ couples to a Bose-Einstein
condensate (BEC). A direct relation between the dephasing of the
internal states of the AQD and the temporal phase fluctuations of
the BEC is established. Based on this relation we suggest a scheme
to probe BEC phase fluctuations nondestructively via dephasing
measurements of the AQD. In particular, the scheme allows to trace
the dependence of the phase fluctuations on the trapping geometry
of the BEC.
\end{abstract}

\submitto{\NJP}

\maketitle

\section{Introduction}
The coherence properties of Bose-Einstein condensates (BECs) have
attracted considerable theoretical and experimental interest since
the first experimental realisation of BECs in trapped ultracold
clouds of alkali atoms \cite{anderson,davis}. Part of this
interest is due to the importance of coherence effects for the
conceptual understanding of BECs and their use as a source of
coherent matter waves. In particular, the absence of spatial
coherence in low dimensional BECs, which exhibit strong spatial
and temporal phase fluctuations, has been investigated
theoretically \cite{petrov2,2dpaper,stoof} and demonstrated for
one dimensional condensates \cite{dettmer,richard}. Moreover,
temporal first and second order phase coherence of an atom laser
beam extracted from a BEC has been observed \cite{kohl,kohl2}.
However, to our knowledge, temporal phase fluctuations in a BEC
have not yet been measured directly, despite the fact that they
ultimately limit the coherence time of an atom laser beam.

In this paper, we propose a scheme to measure temporal phase
fluctuations of the BEC based on a single trapped impurity atom
coupled to the BEC, hereafter called an \emph{atomic quantum dot}
(AQD) \cite{recati}. More specifically, we consider an AQD with
two internal states $\ket{0}$ and $\ket{1}$, where we assume for
simplicity that only state $\ket{1}$ undergoes collisional (s-wave
scattering) interactions with the BEC. This setup could be
implemented using spin-dependent optical potentials \cite{mandel},
where the impurity atom, trapped separately
\cite{diener,meyrath,chuu}, and the BEC atoms correspond to
different internal atomic states. The collisional properties of
the AQD can, to a good approximation, be engineered by means of
optical Feshbach resonances \cite{theis}, which allow to change
the atomic scattering length \emph{locally} over a wide range and
are available even when no magnetic Feshbach resonance exists.

By identifying the combined system of the AQD and the BEC with an
exactly solvable \emph{independent boson model}
\cite{unruh,palma,mahan}, we show that the dephasing of the
internal states due to the asymmetric interaction with the BEC is
directly related to the temporal phase fluctuations. This
dephasing can be detected under reasonable experimental
conditions, for example, in a Ramsey type experiment \cite{bloch},
and hence it is possible to use the AQD to probe BEC phase
fluctuations. Since the phase fluctuations depend strongly on the
temperature and the density of states of the BEC, determined by
the trapping geometry, the proposed scheme allows us to measure
the BEC temperature and to observe the crossover between different
effective BEC dimensions, notably transitions from 3D to lower
dimensions. Our scheme is nondestructive and hence it is possible,
in principle, to investigate the dependence of phase fluctuations
on the BEC dimension and temperature for a single copy of a BEC.

Probing a BEC with an AQD was proposed recently in \cite{recati}
and in a different context in \cite{fischer}. However, in
\cite{recati} two states corresponding to the presence of a single
atom in the trap or its absence were considered, as opposed to
internal atom states. More importantly, in addition to the
collisional interactions the impurity atom was coupled to the BEC
via a Raman transition, allowing the realisation of an independent
boson model with \emph{tunable} coupling. In particular, Recati
\etal \cite{recati} proposed to measure the Luttinger liquid
parameter $K$ by observing the dynamics of the independent boson
model for different coupling strengths.

Other schemes based on interactions between impurity atoms and
BECs have been proposed recently. In \cite{micheli}, for example,
an impurity was used to implement a single atom transistor,
whereas in \cite{klein} a quantum gate exploiting phonon-mediated
interactions between two impurities was investigated. Single atom
cooling in a BEC was considered in \cite{daley} where some aspects
of dephasing of a two state system (qubit) due to BEC fluctuations
were addressed. Whereas their treatment focused on a
three-dimensional BEC and was based on a master equation approach,
in our work we use an analytical approach similar to dephasing
calculations for semiconductor quantum dots \cite{krummheuer,
pazy}.

The paper is organized as follows. In section 2 we introduce our
model for the AQD coupled to the BEC and identify it with an
exactly solvable independent boson model, which enables us to
establish a relation between the temporal phase fluctuations of
the BEC and the dephasing of the AQD. We first consider an
effective $D$-dimensional BEC, which is strongly confined in
$(3-D)$ directions and, as an ideal case, assumed to be
homogeneous along the loosely confined directions. Later we
discuss corrections to this zero potential approximation due to a
shallow trap potential. In section 3 we show by means of a
concrete example how dephasing depends on the interaction time
$\tau$ between the AQD and the BEC, the condensate temperature $T$
and the effective condensate dimension $D$. Section 4 addresses
corrections to our results due to an imperfect measurement of the
AQD state. We conclude in section 5.

\section{The model}
We consider the system of an AQD coupled to a BEC in thermal
equilibrium with temperature $T$. The BEC is confined in a
harmonic trapping potential
\begin{equation}\label{trap}
    V_{trap}(\bv{r})=\frac{m}{2}\Big(\omega_x^2\,x^2+\omega_y^2\,
    y^2+\omega_z^2\,
    z^2\Big)\,,
\end{equation}
with $\bv{r}=(x,y,z)$ the position vector and $m$ the mass of the
condensate atoms. The trap frequencies $\omega_x$, $\omega_y$ and
$\omega_z$ assume the value(s) $\omega_\perp$ or $\omega$, where
$\omega_\perp\gg\omega$ is the trap frequency in the strongly
confined (transverse) direction(s) and $\omega$ the trap frequency
in the loosely confined direction(s). Depending on the number of
strongly confined directions -- none, one or two -- the BEC is
either 3D or assumes effective 2D or 1D character, provided that
thermal excitations are suppressed, i.e. $k_B
T\ll\hbar\omega_\perp$, and that the interaction energy of the
weakly interacting BEC does not exceed the transverse energy, i.e.
$m c^2\ll\hbar\omega_\perp$, with $k_B$ Boltzmann's constant and
$c$ the speed of sound. In directions that are loosly confined the
extension of the trap potential is assumed to be much larger than
the length scale $\sigma$ set by the AQD size, so that it is
justified to approximate the potential by zero, as discussed at
the end of this section. We consider the case where the spectrum
of the BEC excitations, as compared to the impurity spectrum, is
practically continuous.

The AQD consists of an impurity atom in the ground state of a
harmonic trap potential centered at $\bv{r}_0$ and is described by
the wave function $\psi_{\sigma}(\bv{r}-\bv{r}_0)$. We assume that
$\psi_{\sigma}(\bv{r}-\bv{r}_0)$ takes the form of the BEC density
profile in the strongly confined direction(s) and has ground state
size $\sigma$ in the loosely confined direction(s). For example in
case of a 1D condensate
\begin{equation}\label{wavefunction}
    \psi_{\sigma}(\bv{r}-\bv{r}_0)\propto\frac{1}{\sqrt{a_{\perp}^2\,\sigma}}\exp\Big[-\Big(\frac{x-x_{0}}{\sqrt{2}a_{\perp}}\Big)^2-\Big(\frac{y-y_{0}}{\sqrt{2}a_{\perp}}\Big)^2-\Big(\frac{z-z_{0}}{\sqrt{2}\sigma}\Big)^2\Big]\,,
\end{equation}
with the harmonic oscillator length $a_{\perp}\ll\sigma$. We
further assume that the impurity atom has two internal states
$\ket{0}$ and $\ket{1}$, and undergoes s-wave scattering
interactions with the BEC atoms only in state $\ket{1}$.

\subsection{Dephasing in the zero potential approximation}

The total Hamiltonian of the system can be written as
\begin{equation}\label{total}
    H_{tot}=H_{A}+H_{B}+H_{I}\,,
\end{equation}
where $H_{A}=\hbar\Omega\on$, with level splitting $\hbar\Omega$,
is the Hamiltonian for the AQD, $H_{B}$ is the Hamiltonian for the
BEC, and $H_{I}$ describes the interaction between the AQD and the
BEC. Provided that $\sigma/l\gg1$, with $l$ the average
interparticle distance in the BEC, we can represent the
$D$-dimensional (quasi)condensate in terms of the phase operator
$\pp\x$ and number density operator $\hat{n}_{tot}\x=n_0+\nn\x$.
Here, $n_0=l^{-D}$ is the equilibrium density of the BEC, $\nn\x$
is the density fluctuation operator and $\bx$ a $D$-dimensional
vector describing the position in the loosely confined
direction(s). We describe the dynamics of the BEC by a low-energy
effective Hamiltonian \cite{landau,wen}
\begin{equation}\label{sfh1}
    H_{B}=\frac{1}{2}\int\dx\bigg(\frac{\hbar^2}{m}n_0(\nabla\pp)^2(\bv{x})+g\nn^2(\bv{x})\bigg)\,,
\end{equation}
where the interaction coupling constant is defined by $g\equiv
mc^2/n_0$ \footnote{We neglect the weak dependence of $c$ on the
dimensionality of the BEC \cite{zaremba} for simplicity.}. We
shall see that the use of this model is fully justified, even
though it exhibits only Bogoliubov excitations with a linear
dispersion relation $\omega_k=ck$, henceforth called phonons. The
canonically conjugate field operators $\pp\x$ and $\nn\x$ can be
expanded as plane waves
\begin{eqnarray}\label{expand1}
    \pp\x=\vol{1}\sumk\ak^{-1}(\an\pex + \hc)\,,\\\label{expand2}
    \nn\x=\vol{\rmi}\sumk\ak(\an\pex - \hc)\,,
\end{eqnarray}
where $\cre$ and $\an$ are bosonic phonon creation and
annihilation operators and $L^D$ is the sample size. With the
amplitudes $\ak=\sqrt{\hbar\wk/g}$ the Hamiltonian \eref{sfh1}
takes the familiar form
\begin{equation}\label{sfh2}
    H_{B}=\sumk\hbar\wk(\cre\an+\frac{1}{2})\,.
\end{equation}
The coupling between AQD and the BEC occurs in the form of a
density-density interaction
\begin{equation}\label{hint}
    H_{I}=\kappa\,\on\int\dx\,|\bar{\psi}_{\sigma}(\bx-\bx_0)|^2\,\hat{n}_{tot}\x\,,
\end{equation}
where $\bar{\psi}_{\sigma}(\bx-\bx_0)$ is the AQD wave function
integrated over the transverse direction(s), and $\kappa$ the
coupling constant, which can be positive or negative. To avoid
notable deviations of the BEC density from $n_0$ in the vicinity
of the AQD we require $|\kappa|\sim g$ \cite{busch,kalas}. The
interaction Hamiltonian $H_{I}$ acts on the relative phase of the
two internal states, but does not change their population. This
effect is customarily called pure dephasing.

The total Hamiltonian can be identified with an independent boson
model, which is known to have an exact analytic solution
\cite{unruh,palma,mahan}. Inserting the explicit expression
\eref{expand2} for $\nn\x$ into equation \eref{hint} we can
rewrite the total Hamiltonian as
\begin{equation}\label{spinboson}
    \fl H_{tot}=(\kappa
    n_{0}+\hbar\Omega)\on+\sumk\Big(g_{\bv{k}}\cre+g_{\bv{k}}^{*}\an\Big)\on+\sumk\hbar\wk(\cre\an+\frac{1}{2})\,,
\end{equation}
where $\kappa n_{0}$ is the mean field shift. The coupling
coefficients $g_{\bv{k}}$, which contain the specific
characteristics of the system, are given by
\begin{equation}\label{g}
    g_{\bv{k}}=-\vol{\rmi\kappa}\,\ak f_{\bv{k}}\,,
\end{equation}
with the Fourier transform of the AQD density
\begin{equation}\label{formfactor}
    f_{\bv{k}}=\int\dx\,|\,\bar{\psi}_{\sigma}(\bx-\bx_0)|^2\nex\,.
\end{equation}
As can be seen from expression \eref{spinboson} the effect of the
AQD in state $\ket{1}$ is to give rise to a displaced harmonic
oscillator Hamiltonian for each phonon mode. This problem is
identical with the problem of a charged harmonic oscillator in a
uniform electric field and can be solved by shifting each phonon
mode to its new equilibrium position \cite{mahan}. As a
consequence the AQD does not change the phonon frequencies and
hence the phase fluctuations are the same in the presence of the
AQD as in its absence.

The state of the system is described by the density matrix
$\rho_{tot}(t)$. We assume that the AQD is in state $\ket{0}$ for
$t< 0$ and in a superposition of $\ket{0}$ and $\ket{1}$ at $t=0$,
which can be achieved, for example, by applying a short laser
pulse. Hence the density matrix $\rho_{tot}(t)$ at time $t=0$ is
of the form
\begin{equation}\label{density}
    \rho_{tot}(0)=\rho(0)\otimes\rho_{B}\,,\qquad\rho_{B}=\frac{1}{Z_{B}}\exp[-H_{B}/(k_B
    T)]\,,
\end{equation}
where $\rho(0)$ is the density matrix for the AQD and $\rho_{B}$
the density matrix of the BEC in thermal equilibrium, with $Z_{B}$
the BEC partition function. After a change to the interaction
picture the time evolution operator of the total system takes the
form \cite{palma}
\begin{eqnarray}\label{evolution}
    U(t)
    &=&\exp\bigg[\on\sumk\Big(\beta_{\bv{k}}\cre-\beta_{\bv{k}}^{*}\an\Big)\bigg]\,,
\end{eqnarray}
with $\beta_{\bv{k}}=g_{\bv{k}}(1-\exp[\rmi\wk t])/(\hbar\wk)$.
The reduced density matrix of the AQD can be determined from the
relation $\rho(t)=\Tr_{B}\{U(t)\rho_{tot}(0)U^{-1}(t)\}$, where
$\Tr_{B}$ is the trace over the BEC. The coherence properties of
the AQD are governed by the off-diagonal matrix elements
\begin{eqnarray}\label{relation}
    \rho_{10}(t)=\rho_{01}^{*}(t)=\rho_{10}(0)\rme^{-\gamma(t)}\,,
\end{eqnarray}
with the dephasing function
\begin{equation}\label{dephase}
    \gamma(t)=-\ln\bigg\langle\exp\bigg[\sumk\Big(\beta_{\bv{k}}\cre-\beta_{\bv{k}}^{*}\an\Big)\bigg]\bigg\rangle\,,
\end{equation}
where angular brackets denote the expectation value with respect
to the thermal distribution $\rho_{B}$.

The dephasing function $\gamma(t)$ can be expressed in terms of
the phase operator in the interaction picture $\pp\xt=\exp[\rmi
H_{B} t/\hbar]\pp\x\exp[-\rmi H_{B} t/\hbar]$. We introduce the
coarse-grained phase operator averaged over the AQD size $\sigma$
\begin{equation}\label{cg}
    \ppc\xot\equiv\int\dx|\bar{\psi}_{\sigma}(\bx-\bx_0)|^2\pp\xt
\end{equation} and the phase difference
$\delta\ppc\xot\equiv\ppc\xot-\ppc(\bx_0,0)$ to rewrite the phase
coherence as
\begin{eqnarray}\label{crucial}
      \rme^{-\gamma(t)}&=\Big\langle\exp\Big[\,\rmi\frac{\kappa}{g}\,\delta\ppc\xot\Big]\Big\rangle\\\label{even}
      &=\exp\left[-\frac{1}{2}\,\Big(\frac{\kappa}{g}\Big)^2\,\Big\langle(\delta\ppc)^2\xot\Big\rangle\right]\,,
\end{eqnarray}
where the second equality can be proven by direct expansion of the
exponentials \cite{mahan}. Thus we have established a direct
relation between the dephasing of the AQD and the temporal phase
fluctuations of the BEC averaged over the AQD size $\sigma$.
Moreover, the phase coherence \eref{even} is closely related to
the correlation function
$\langle\Psi^{\dag}(\bv{x},t)\Psi(\bv{x}^{\prime},t^{\prime})\rangle$,
with $\Psi(\bv{x},t)$ the bosonic field operator describing the
BEC in the Heisenberg representation. In the long-wave
approximation, where the density fluctuations are highly
suppressed, we have \cite{landau,wen}
\begin{equation}\label{corrr}
\langle\Psi^{\dag}(\bv{x}_0,t)\Psi(\bv{x}_0,0)\rangle\approx
n_0\exp\left[-\frac{1}{2}\,\Big\langle(\delta\pp)^2\xot\Big\rangle\right]\,.
\end{equation}
Therefore expression \eref{even} can be interpreted as a smoothed
out temporal correlation function of the BEC field operator
$\Psi(\bv{x},t)$ provided that $\kappa=g$.

To further investigate the effect of the BEC phase fluctuations on
the AQD we require an explicit expression for the dephasing
function $\gamma(t)$ depending on the parameters of our system.
For this purpose we express the dephasing function $\gamma(t)$ in
terms of the coupling coefficients $g_{\bv{k}}$ \cite{palma}
\begin{equation}\label{gamma}
    \gamma(t)=\sumk|g_{\bv{k}}|^2\coth\Big(\frac{\hbar\wk}{2k_{B}T}\Big)\frac{1-\cos(\wk
    t)}{(\hbar\wk)^2}
\end{equation}
and take the thermodynamic limit of $\gamma(t)$, which amounts to
the replacement $\sumk\rightarrow \int_0^\infty\rmd k\,g(k)$, with
the density of states $g(k)=S_D\,L^D k^{D-1}$, $S_D\equiv
D/[2^D\pi^{D/2}\Gamma(D/2+1)]$ and $\Gamma(x)$ the gamma function.
Substituting expression \eref{g} for $g_{\bv{k}}$ we find
\begin{equation}\label{gammaint}
    \gamma(t)=\frac{S_D}{2}\frac{\kappa^2}{g}\int_0^\infty
    \rmd k\,k^{D-1}|f_{\bv{k}}|^2\coth\Big(\frac{\hbar\wk}{2k_{B}T}\Big)\frac{1-\cos(\wk
    t)}{\hbar\wk}\,.
\end{equation}
The integral is well defined due to the factor
$|f_{\bv{k}}|^2=\exp[-\sigma^2 k^2/2]$, which provides a natural
upper cut-off. Since for typical experimental parameters
$\xi/l\sim1$, with $\xi\sim\hbar/(mc)$ the healing length, the
condition $\sigma/l\gg1$ implies that $\sigma\gg\xi$. Thus the
upper cut-off at $k\sim1/\sigma$ is still in the phonon regime,
which justifies the use of the effective Hamiltonian \eref{sfh1}.

\subsection{Corrections due to a shallow potential}

We now give the conditions under which the effects of a shallow trap
on result \eref{gammaint} are negligible in the thermodynamic limit.
This  limit is taken such that $\omega\propto 1/R$, with $R$ the
radius of the BEC, to ensure that the BEC density at the center of
the trap remains finite. To determine the trap-induced corrections
to the density of states and the phonon wave functions, which affect
$f_{\bv{k}}$ defined by \eref{formfactor}, we use a semi-classical
approach based on the classical Hamiltonian \cite{london}
\begin{equation}\label{claha}
    H(\bv{p},\bv{x})=c(\bv{x})|\bv{p}|+V(\bv{x})\,,
\end{equation}
with $c(\bv{x})=c\,(1-\bv{x}^2/R^2)^{1/2}$ the position dependent
speed of sound, $\bv{p}$ the phonon momentum and $V(\bv{x})=m
\omega^2 \bv{x}^2/2$ the shallow trap potential. The range of
phonon energies $\varepsilon$ relevant for the AQD dephasing is
$\hbar\omega/2<\varepsilon<\hbar c/\sigma$, where the lower bound
goes to zero in the thermodynamic limit.

Given the semi-classical phonon wave functions, we find that
corrections to $f_{\bv{k}}$ are negligible if $\sigma\ll
L_{\varepsilon}$, with
$L_\varepsilon=\sqrt{2\varepsilon/(m\omega^2)}$ the classical
harmonic oscillator amplitude. For the density of states
$g(\varepsilon)$ we use the semi-classical expression
\begin{equation}\label{dos}
    g(\varepsilon)=\frac{1}{(2\pi\hbar)^D}\int\dx
    \int\rmd^{D}\textbf{p}\;
    \delta\Big(\varepsilon-H(\bv{p},\bv{x})\Big)
\end{equation}
to obtain $g(\varepsilon)\propto
L_\varepsilon^{D}\varepsilon^{D-1}\,[1+\mathcal{O}(L_\varepsilon/R)]$.
Thus in the regime where $\sigma\ll L_\varepsilon\ll R$ and after
the substitutions $L\rightarrow L_\varepsilon$ and
$\hbar\wk\rightarrow\varepsilon$ we have
\begin{equation}\label{re}
     \gamma(t)\propto\frac{\kappa^2}{g}\int_0^\infty
    \rmd
    \varepsilon\,\varepsilon^{D-1}|f_{\varepsilon}|^2\coth\Big(\frac{\varepsilon}{2k_{B}T}\Big)\frac{1-\cos(\varepsilon
    t/\hbar)}{\varepsilon}\,,
\end{equation}
which up to numerical constants is identical to expression
\eref{gammaint}. The conditions on $L_\varepsilon$, together with
the bounds of the phonon spectrum, imply that the dephasing in a
shallow trap does not differ from the homogeneous case if
$\xi\ll\sigma\ll a_{\omega}$, with
$a_{\omega}=\sqrt{\hbar/(m\omega)}$ the harmonic oscillator
length.

\section{Application}

\begin{figure}[h]
\begin{center}
\begin{tabular}{c}
\epsfbox{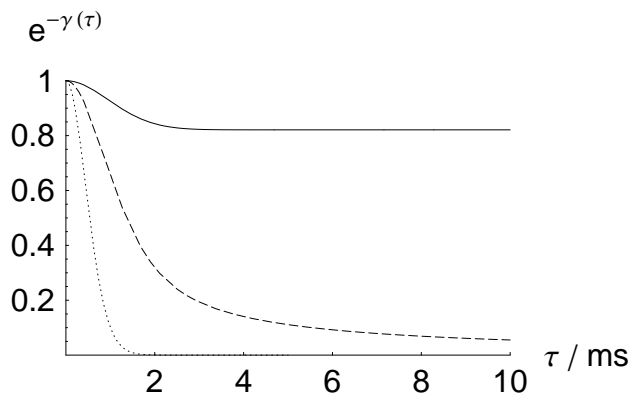}
\end{tabular}
\end{center}
\caption{\label{time} The phase coherence $\rme^{-\gamma(\tau)}$
as a function of the interaction time $\tau$. In the regime
$\tau\gg\sigma/c$ the phase coherence approaches a non-zero value
in 3D (solid), falls off as a power law in 2D (dashed) and shows
exponential decay in 1D (dotted). The system parameters were set
to $m=10^{-25}\mathrm{kg}$, $l=5\times10^{-7}\mathrm{m}$,
$c=10^{-3}\mathrm{m s}^{-1}$, $T=2\times10^{-7}\mathrm{K}$,
$\sigma=10^{-6}\mathrm{m}$ and $\kappa=g$.}
\end{figure}

In this section we discuss by means of a concrete example
\footnote{For the numerical values of the system parameters see
the caption of \fref{time}.} how the AQD dephasing depends on the
interaction time $\tau$, the condensate temperature $T$ and the
effective condensate dimension $D$ \footnote{A discussion of a
similar model in the context of quantum information processing can
be found in \cite{palma}.}. To find the phase coherence
$\rme^{-\gamma(\tau)}$ we have evaluated the integral
\eref{gammaint} numerically (shown in figures 1, 2 and 3) and
analytically in the high temperature regime $k_{B}T\gg\hbar
c/\sigma$.

\Fref{time} shows the phase coherence $\rme^{-\gamma(\tau)}$ as
function of the interaction time $\tau$. The evolution of the
phase coherence is split into two regimes separated by the typical
dephasing time $\sigma/c\sim 10^{-3}$s, which can be viewed as the
time it takes a phonon to pass through the AQD. It follows from
the analytic calculations that the asymptotic behaviour of
$\rme^{-\gamma(\tau)}$ for times $\tau\gg\sigma/c$ is
\begin{equation}\label{coor1}
   \rme^{-\gamma(\tau)}=\cases{C_{T}\exp\left[-\Big(\frac{\kappa}{g}\Big)^2\,\frac{mc\,k_{B}T}{2\hbar^2 n_0}\,\tau\right]&for $D=1$\\
   C_{T}^{\prime}\Big(\frac{\sigma}{c\tau}\Big)^\nu&for $D=2$\\
    \exp\left[-\Big(\frac{\kappa}{g}\Big)^2\frac{m\,k_{B}T}{(2\pi)^{3/2}\hbar^2 n_0\sigma}\right]&for $D=3$}
\end{equation}
where $\nu=(\kappa/g)^2 m k_{B}T/(2\pi\hbar^2 n_0)$ and $C_{T}$,
$C_{T}^{\prime}$ are temperature dependent constants. Thus the
phase coherence tends asymptotically to a non-zero value in 3D,
falls off as a power law in 2D and shows exponential decay in 1D.
This result reflects the fact that the physics of 1D and 2D
condensates differs significantly from that of a 3D condensate.
Given the relation between the phase coherence and the BEC
correlations, discussed in section 2.1, we note that the results
for the phase coherence are consistent with findings for temporal
and spatial coherence properties of the BEC \cite{landau,wen},
where in the latter case the distance $|\bv{x}|$ has to be
identified with $c\tau$.

In particular in 3D the dephasing of the AQD is incomplete under
reasonable experimental conditions due to the suppressed influence
of low frequency fluctuations, which has been noted in the context
of quantum information processing \cite{palma,daley}. However, the
residual coherence of the AQD will eventually disappear due to
processes as, for example, inelastic phonon scattering
\cite{daley}, which are not taken into account in our model. We
also point out that, in contrast to the remnant coherence of the
AQD, the typical dephasing time does not depend on the coupling
constant $\kappa$.

\begin{figure}[h]
\begin{center}
\begin{tabular}{c}
\epsfbox{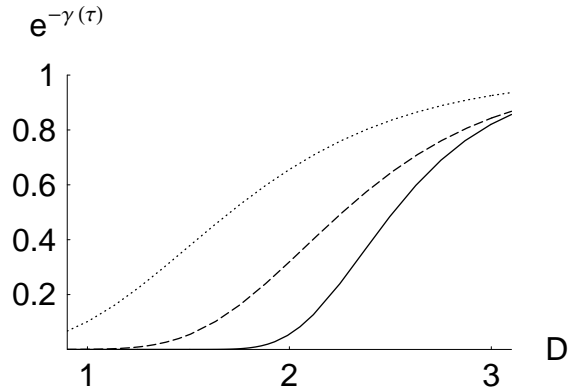}
\end{tabular}
\end{center}
\caption{\label{dime}The phase coherence $\rme^{-\gamma(\tau)}$ as
a function of the effective condensate dimension $D$ for the
interaction times $\tau_1=\sigma/c$ (dotted), $\tau_2=2\,\sigma/c$
(dashed) and $\tau_3=10\,\sigma/c$ (solid). The phase coherence
drops significantly while the BEC excitations in the strongly
confined directions are frozen out. The plot was produced with the
same set of parameters as in \fref{time}.}
\end{figure}

The fact that phase fluctuations depend strongly on the density of
states, or equivalently the dimension of the BEC, allows us to
observe the crossover between different effective dimensions,
especially transitions from 3D to lower dimensions. \Fref{dime}
shows the phase coherence $\rme^{-\gamma(\tau)}$ as a function of
the dimension $D$ for the interaction times $\tau_1=\sigma/c$,
$\tau_2=2\,\sigma/c$ and $\tau_3=10\,\sigma/c$. The phase
coherence $\rme^{-\gamma(\tau)}$ drops significantly as the BEC
excitations in the strongly confined direction(s) are frozen out.
Therefore we expect that the change of phase fluctuations
depending on the effective dimension should be experimentally
observable.

\begin{figure}[h]
\begin{center}
\begin{tabular}{cc}
   \raisebox{4cm}[-4cm]{1D} & \epsfbox{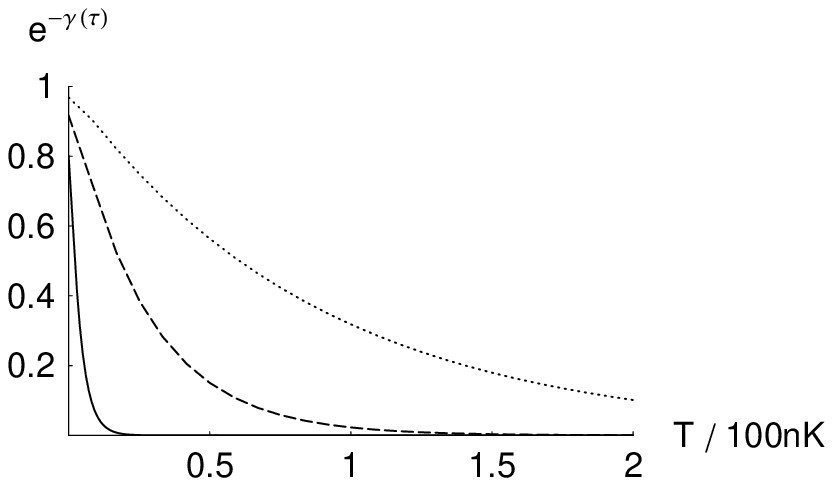} \\
   \raisebox{4cm}[-4cm]{2D} & \epsfbox{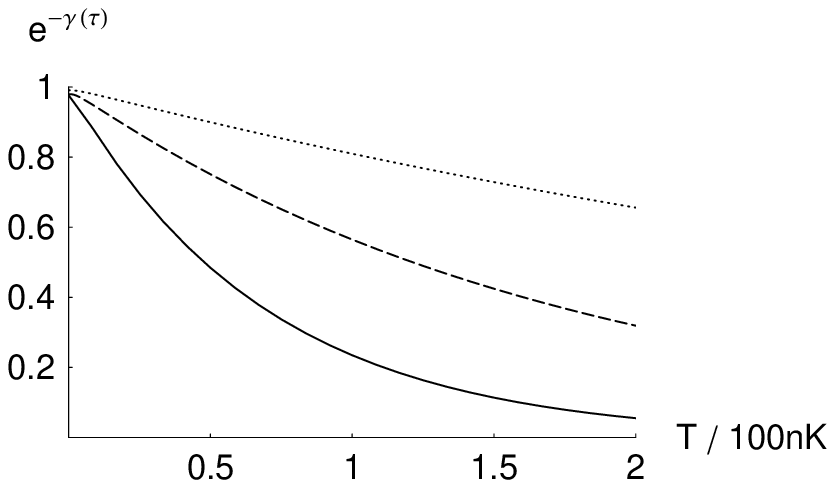} \\
   \raisebox{4cm}[-4cm]{3D} & \epsfbox{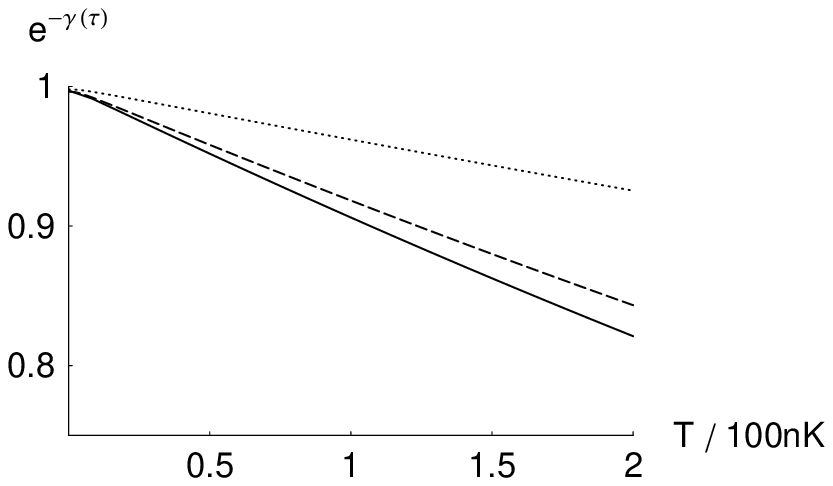} \\
\end{tabular}
\end{center}
\caption{\label{temp}The phase coherence $\rme^{-\gamma(\tau)}$ as
a function of temperature $T$ for interaction times
$\tau_1=\sigma/c$ (dotted), $\tau_2=2\,\sigma/c$ (dashed) and
$\tau_3=10\,\sigma/c$ (solid). In 1D and 2D the interaction time
$\tau$ can be chosen to assure that dephasing changes
significantly with temperature, whereas in 3D dephasing is
independent of $\tau$ for $\tau\gg\sigma/c$. The plot was produced
with the same set of parameters as in \fref{time}, except for the
temperature.}
\end{figure}

In addition, the AQD can be used to measure the BEC temperature
since the dephasing function $\gamma(\tau)$ is approximately
proportional to the temperature $T$ according to equation
\eref{coor1}. \Fref{temp} shows the relation between the phase
coherence $\rme^{-\gamma(\tau)}$ and the BEC temperature $T$ in
all three dimensions for the interaction times $\tau_1=\sigma/c$,
$\tau_2=2\,\sigma/c$ and $\tau_3=10\,\sigma/c$. In 1D and 2D the
interaction time $\tau$ can be chosen to assure that the AQD
dephasing changes significantly with temperature, whereas in 3D
the interaction time $\tau$ has little influence on the thermal
sensitivity of the AQD.

In the high temperature regime $k_{B}T\gg\hbar c/\sigma$ the phase
coherence is reduced mainly due to thermal phonons. However, even
at zero temperature coherence is lost due to purely quantum
fluctuations. This loss of coherence takes place on the same time
scale as in the high temperature regime, but is incomplete in both
3D and 2D, which can be seen by evaluating the integral
\eref{gammaint} in the limit $t\gg\sigma/c$ with $T=0$.

\section{Measurement of the internal states}

The AQD dephasing can be detected in a Ramsey type experiment
\cite{bloch}: The AQD is prepared in state $\ket{0}$ and hence
initially decoupled from the BEC. A first $\pi/2$-pulse at $t=0$
changes the state to a superposition $(\ket{0}+\ket{1})/\sqrt{2}$
and a spin-echo-type $\pi$-pulse at $t=\tau/2$ neutralizes the
mean field shift. After a second $\pi/2$-pulse at $t=\tau$ the AQD
is found in state $\ket{1}$ with probability
\begin{equation}\label{proba1}
    P(\ket{1})=\frac{1}{2}(1-\rme^{-\gamma(\tau)})\,.
\end{equation}

However, this result is altered by decay of state $\ket{1}$ into
state $\ket{0}$, atom loss, imperfect (noisy) detection of state
$\ket{1}$, and additional dephasing due to environmental noise. We
subsume these processes into three phenomenological constants,
namely the detection probability $P_d$, the probability of a
spurious detection $P_s$, and the dephasing rate $\gamma_d$, which
can all be determined experimentally. Taking these effects into
account we find an effective probability $\tilde{P}$ to detect
state $\ket{1}$
\begin{equation}\label{peff}
    \tilde{P}(\ket{1})=\frac{1}{2}P_d\,(1-\rme^{-\gamma(\tau)-\gamma_d\tau})+P_s
\end{equation}
and the visibility
$V\equiv(\tilde{P}_{max}-\tilde{P}_{min})/(\tilde{P}_{max}+\tilde{P}_{min})$
in the limit $\gamma_d\,\tau\ll1$ is given by
\begin{equation}\label{visibility}
    V=\frac{1-\gamma_d\,\tau}{1+\gamma_d\,\tau+4\,P_{s}/P_{d}}\,,
\end{equation}
where $\tilde{P}_{max}$ and $\tilde{P}_{min}$ correspond
respectively to $\tilde{P}(\ket{1})$ in the case of complete
dephasing and no dephasing due to phase fluctuations. Kuhr \etal
\cite{kuhr} recently demonstrated state-selective preparation and
detection of the atomic hyperfine state for single cesium atoms
stored in a red-detuned dipole trap. They showed that dephasing
times of 146ms and ratios $P_{s}/P_{d}$ of the order of
$5\times10^{-2}$ are achievable. If we choose
$\gamma_d=10\mathrm{s}^{-1}$, $\tau=10\mathrm{ms}$ and
$P_{s}/P_{d}=5\times10^{-2}$ we find a visibility of $V=69\%$,
which shows that our scheme is feasible even in the presence of
additional dephasing due to environmental noise.

\section{Conclusion}
We have shown that the dephasing of the internal states of an AQD
coupled to a BEC is directly related to the temporal phase
fluctuations of the BEC. Based on this relation we have suggested
a scheme to probe BEC phase fluctuations nondestructively via
measurements of the AQD coherences, for example, in a Ramsey type
experiment. It was shown that the scheme works for a BEC with
reasonable experimental parameters even in the presence of
additional dephasing due to environmental noise. In particular,
the scheme allows us to trace the dependence of the phase
fluctuations on the trapping geometry of the BEC and to measure
the BEC temperature.

Our scheme is applicable even if the BEC is trapped in not
strongly confined directions provided that the trapping potential
is sufficiently shallow. We expect that the observed dephasing
will be qualitatively different from our results only if the AQD
size $\sigma$ is comparable to the classical harmonic oscillator
amplitude $L_\varepsilon\sim1/\omega$. In addition, our findings
indicate that the use of AQDs for quantum information processing,
proposed recently in \cite{klein,daley}, may be constrained
because of the unfavorable coherence properties of low dimensional
BECs.

The results in \cite{palma} suggest a natural extension of our
work to entangled states between several AQDs, which might lead to
a probe with higher sensitivity. However, the experimental
requirements for the state-selective preparation and detection of
entangled states are considerably higher than for a single AQD,
which has to be considered in the analysis of an extended scheme.

\ack This work was supported by the EPSRC (UK) through the project
EP/C51933/1. MB thanks Alexander Klein for helpful discussions and
the Berrow Scholarship of Lincoln College (Oxford) for financial
support.

\end{document}